\documentclass[fleqn,usenatbib,useAMS]{mnras}

\usepackage{graphicx} 
\usepackage{amsmath} 
\usepackage{amssymb} 
\usepackage{multicol} 
\usepackage{bm} 
\usepackage{pdflscape} 
\usepackage{graphicx,xcolor,hyperref}

\usepackage[T1]{fontenc}
\usepackage{ae,aecompl}

\newcommand{\euc}{\textit{Euclid}}
\newcommand{\ha}{H$\alpha$}
\newcommand{\hb}{H$\beta$}
\newcommand{\oiii}{O\textsc{iii}}
\newcommand{\oiiihb}{O\textsc{iii}+H$\beta$}
\newcommand{\oii}{O\textsc{ii}}
\newcommand{\nii}{N\textsc{ii}}

\newcommand{\de}{\mathrm{d}}

\newcommand{\be}{\begin{equation}}
\newcommand{\ee}{\end{equation}}
\newcommand{\bea}{\begin{eqnarray}}
\newcommand{\eea}{\end{eqnarray}}
\newcommand{\bfig}{\begin{figure}}
\newcommand{\efig}{\end{figure}}

\title[Extending H$\alpha$ galaxy surveys to higher redshifts]{High-redshift cosmology with oxygen lines from H$\alpha$ surveys}

\author[J.\ Fonseca \& S.\ Camera]{Jos\'e Fonseca$^{1,2}$\thanks{josecarlos.s.fonseca@gmail.com} and Stefano Camera$^{3,4,5}$\thanks{stefano.camera@unito.it}\\
$^1$Dipartimento di Fisica ``G. Galilei'', Universit\`a degli Studi di Padova, Via Marzolo 8, 35131 Padova, Italy\\
$^2$INFN -- Istituto Nazionale di Fisica Nucleare, Sezione di Padova, Via Marzolo 8, 35131 Padova, Italy\\
$^3$Dipartimento di Fisica, Universit\`a degli Studi di Torino, Via P.\ Giuria 1, 10125 Torino, Italy\\
$^4$INFN -- Istituto Nazionale di Fisica Nucleare, Sezione di Torino, Via P.\ Giuria 1, 10125 Torino, Italy\\
$^5$INAF -- Istituto Nazionale di Astrofisica, Osservatorio Astrofisico di Torino, Strada Osservatorio 20, 10025 Pino Torinese, Italy
}

\date{}

\pubyear{2020}

\begin{document}
\label{firstpage}
\pagerange{\pageref{firstpage}--\pageref{lastpage}}
\maketitle

\begin{abstract}

A new generation of cosmological experiments will spectroscopically detect the \ha\ line from ELGs at optical/near-infrared frequencies. Other emission lines will also be present, which may come from the same \ha\ sample or constitute a new galaxy sample altogether. Our goal is to assess the value, for cosmological investigation, of galaxies at $z\gtrsim2$ present in \ha\ galaxy surveys and identifiable by the highly redshifted ultra-violet and optical lines---namely the \oii\ line and the \oiii\ doublet in combination with the \hb\ line. We use state-of-the-art luminosity functions to estimate the number density of \oiiihb\ and \oii\ ELGs. We study the constraining power of these high-redshift galaxy samples on cosmological parameters such as the BAO amplitude, $H(z)$, $D_{\rm A}(z)$, $f\sigma_8(z)$, and $b\sigma_8(z)$ for different survey designs. We present a strong science case for extracting the \oiiihb\ sample, which we consider as an independent probe of the Universe in the redshift range $2-3$. Moreover, we show that the \oii\ sample can be used to measure the BAO and growth of structure above $z=3$; albeit it may be shot-noise dominated, it will nonetheless provide valuable tomographic information. We discuss the scientific potential of a sample of galaxies which, so far, has been mainly considered as a contaminant in \ha\ galaxy surveys. Our findings indicate that planed \ha\ surveys should include the extraction of these oxygen-line samples in their pipeline, to enhance their scientific impact on cosmology.


\end{abstract}

\begin{keywords}
large-scale structure of the Universe, cosmology: miscellaneous
\end{keywords}




\section{Introduction}
\label{sec:intro}


Emission-line galaxies (ELGs), which are mainly star-forming galaxies, have UV and optical prominent lines that we use to determine the redshift of each individual ELG. Such lines include Ly$\alpha$ ($121.6\,\mathrm{nm}$), \oii\ ($372.7\,\mathrm{nm}$ and $372.9\,\mathrm{nm}$), Ne\textsc{iii} ($387.0\,\mathrm{nm}$), \hb\ ($486.1\,\mathrm{nm}$), the \oiii\ doublet ($495.9\,\mathrm{nm}$ and $500.7\,\mathrm{nm}$), O\textsc{i} ($630.0\,\mathrm{nm}$), \nii\ ($654.8\,\mathrm{nm}$ and $658.3\,\mathrm{nm}$), \ha\ ($656.5\,\mathrm{nm}$), S\textsc{ii} ($6717\,\mathrm{nm}$ and $6731\,\mathrm{nm}$), and other weaker lines. \ha\ is the strongest optical emission line from star-forming galaxies, second only to Ly$\alpha$ in the UV, and followed by the oxygen lines \oiii\ and \oii. In practice, \nii\ is nearly indistinguishable from \ha\ and represents only a minor contribution to the signal. It is thus natural to choose \ha\ when devising cosmological surveys targeting ELGs. But the \ha\ line with a rest wavelength of $656.5\,\mathrm{nm}$ is quickly redshifted into the near-infrared where the atmosphere transparency is reduced, drastically diminishing the number of detectable galaxies from the ground. For this reason, future optical and near-infrared surveys will be in space. The three planned surveys are: the Europe-led ESA's flagship mission, the \euc\ satellite\footnote{\url{https://www.euclid-ec.org}} \citep{Laureijs:2011gra}, which will take spectra of millions of ELGs to identify their redshift; the USA-led NASA WFIRST satellite\footnote{\url{https://wfirst.gsfc.nasa.gov}} \citep[Wide Field Infrared Survey Telescope,][]{Spergel:2015sza}; and another NASA mission called SPHEREx\footnote{\url{http://spherex.caltech.edu}} \citep[Spectro-Photometer for the History of the Universe, Epoch of Reionization, and Ices Explorer,][]{Dore:2014cca}, which will complement the previous two.  The design of the satellites has been optimised for a wide range of scientific goals, including several trade-offs between sensitivity, surveyed area, wavelength coverage, available emission lines from ELGs and so on. This has resulted into different sky area coverages and wavelength ranges in the optical and near-infrared bands, with some overlap among them, which we summarise in \autoref{fig:z_vs_lamb}. In Table \ref{tab:exp_details} we summarise the wavelength coverage and spectral resolution of the spectroscopic specifications of these experiments.

\bfig
\centering
\includegraphics[width=\columnwidth]{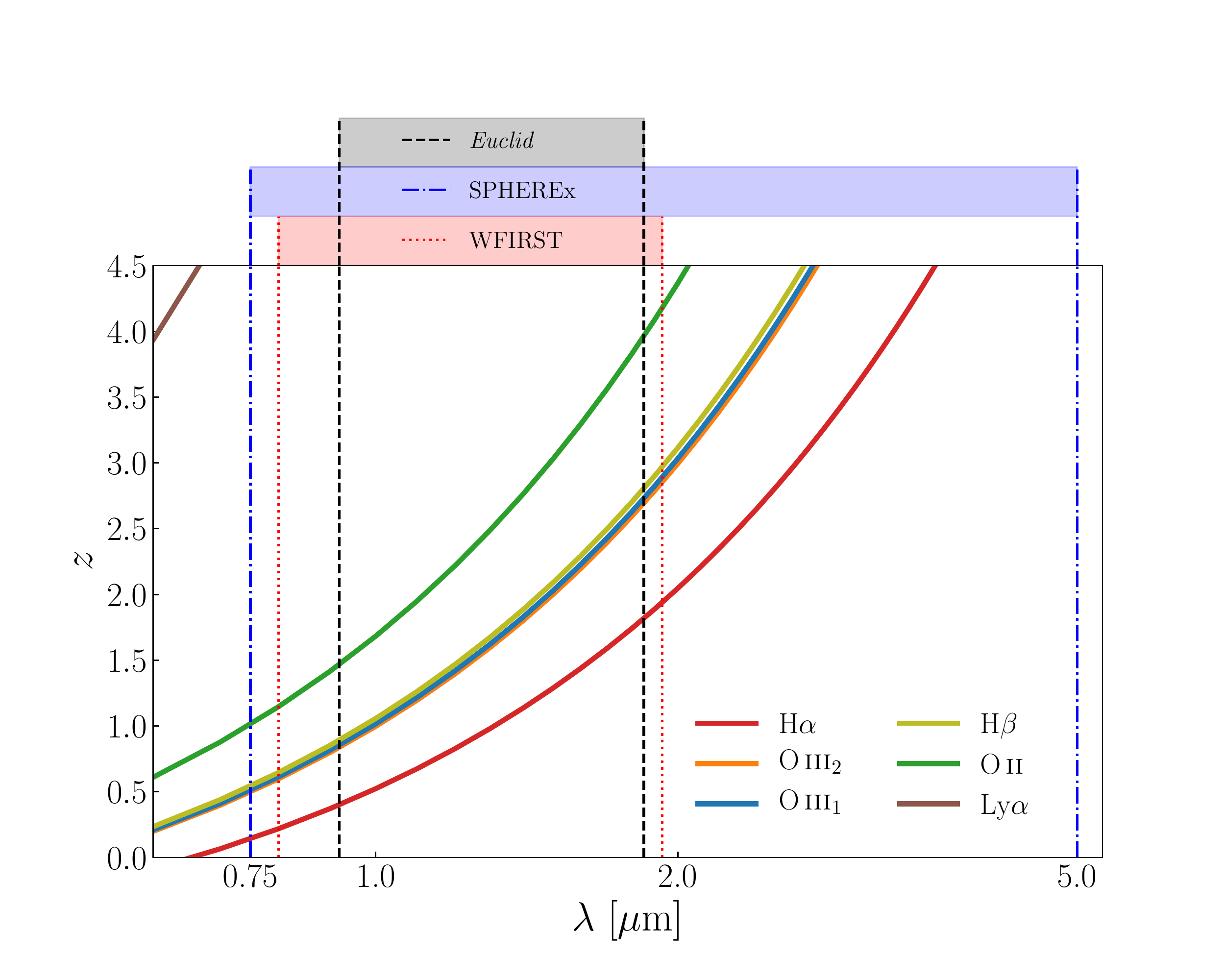}
\caption{Redshift of different emission lines as a function of the observed wavelength. Vertical lines indicate the wavelength coverage of the spectrographic instruments of different experiments: \euc\ (black, dashed line), SPHEREx (blue, dot-dashed line), and WFIRST (red, dotted line).}
\label{fig:z_vs_lamb}
\efig


\begin{table*}
\centering            
\caption{Summary of the details of spectroscopic part of future space-based NIR galaxy surveys: \euc\ \citep{Blanchard:2019oqi}, WFIRST \citep{Spergel:2015sza} and SPHEREx \citep{Dore:2014cca}. Updated numbers for the experiments were obtained from the respective websites.}
\begin{tabular}{ccccc}
\hline
Survey &  Range [$\mu$m]&$R=\lambda/\Delta\lambda$ & Flux threshold [$\times 10^{-16}$ erg/s/cm$^2$] & Survey Area   \\
\hline\hline
\euc$^1$ & $0.92 - 1.25$ | $1.25 - 1.85$  & $380$     & Wide: 2 | Deep: $\sim0.8$; & $15000\,\deg^2$ | $40\,\deg^2 $\\
 WFIRST$^2$ & $1.0 - 1.93$  | $0.8 - 1.8$ & $450 - 850$ | $70 -140$  & $1.2$ & $2227\,\deg^2$ \\
SPHEREx$^3$ &  $0.75 - 2.42$ | $2.42 - 3.82$ & $41$ | $35$ & $\sim 50$& Full sky\\
&  $3.82 - 4.42$ | $4.42 - 5.00$ & $110$ | $130$ & & \\
\hline
\end{tabular}
\label{tab:exp_details}     
\end{table*}

Despite the prominence of the \ha\ line, other emission lines are used to identify the redshift of ELGs, as it is already done by other ground-based spectroscopic galaxy surveys. This has been the case for past surveys such as SDSS \citep{Strauss:2002dj}, WiggleZ \citep{2008A&G....49e..19B}, GAMA \citep{Baldry_2010}, VIPERS \citep{2018A&A...609A..84S}, and current surveys such as DESI \citep{Aghamousa:2016zmz}. While SPHEREx will always have a complete set of lines to fully determine the redshift of a given galaxy, \euc\ and WFIRST will only have a subset of these lines available (see \autoref{fig:z_vs_lamb}). 

Let us take the example of \euc. Ly$\alpha$ will mainly come from redshifts well inside the epoch of reionisation and we expect it to be sufficiently faint, such that it will not substantially contaminate the sample. But the oxygen lines are strong and high-$z$ ELGs may contaminate the \ha\ sample. Depending on the emitting redshift and experimental resolution, the \oiii\ doublet, and \hb\ will be indistinguishable so we will bundle them together for simplicity. Even if the experiment provides enough wavelength resolution, these lines are close enough to be considered as a distinctive sample that in practice increases the signal-to-noise ratio of detection. Thus, in the observing window of \euc, \ha\ will see ELGs from $z\in [0.40,1.82]$, \oiiihb\ will see them in the range $z\in [0.84,2.81]$, and \oii\ in the interval $z\in [1.47,3.96]$. In Figure \ref{fig:z_lines_exp} we compare the redshift covered by each line in the three experiments we consider. For SPHEREx specifications these lines can reach redshifts during the Epoch of Reionization (EoR), thus we truncated the figure at $z=4.5$.

\bfig
\centering
\includegraphics[width=\columnwidth]{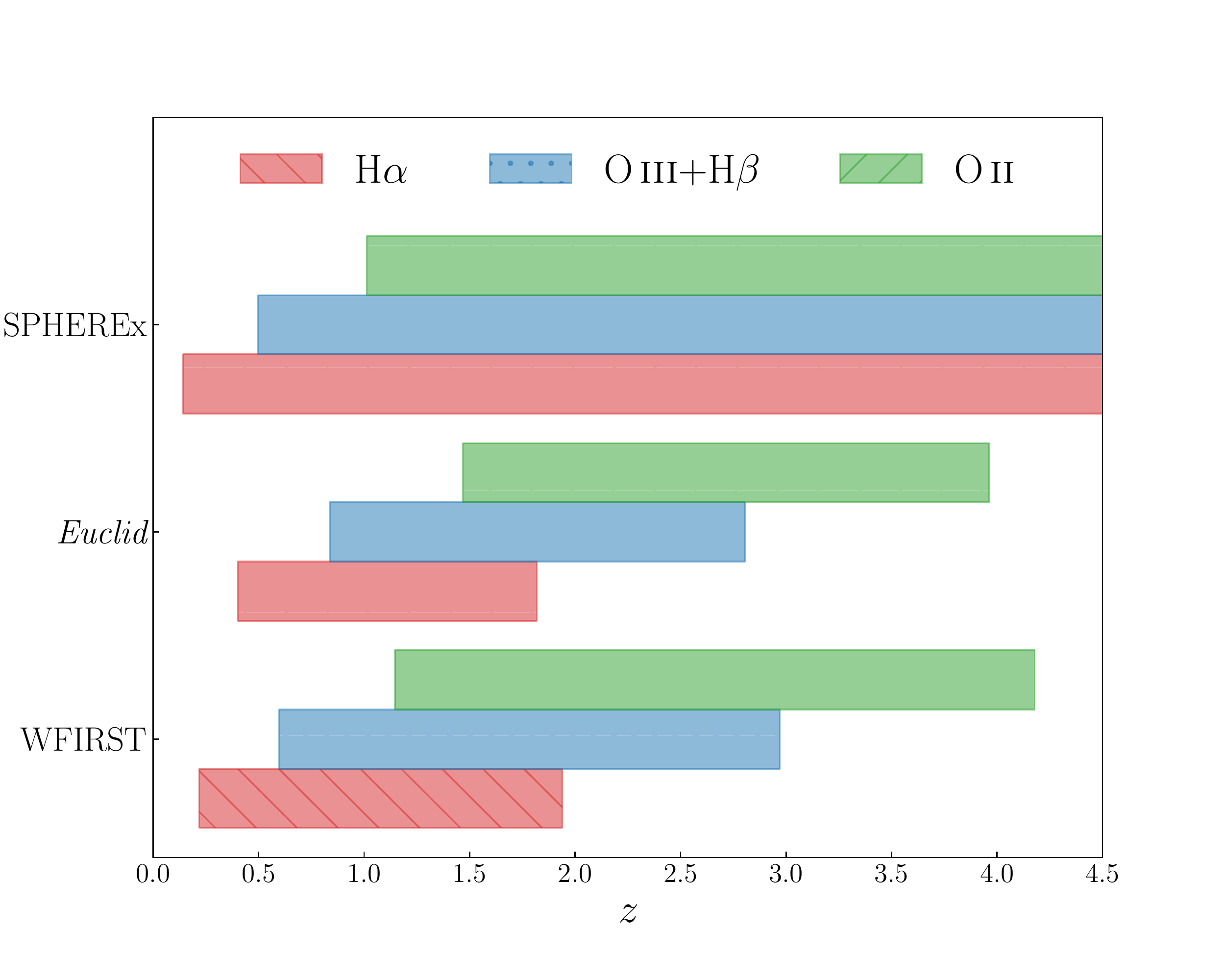}
\caption{Redshift ranges seen by \ha\ (red), \oiiihb\ (blue), and \oii\ (green) for WFIRST (bottom), \euc\ (middle), and SPHEREx (top). Note that the redshift range of SPHEREx extends to well inside the epoch of reionisation, but for better visualisation we truncate the figure at $z=4.5$.}
\label{fig:z_lines_exp}
\efig

Hence, it is clear that for the same wavelength coverage one will observe low-redshift \ha\ emitters as well as high-redshift galaxies identifiable by \oiiihb\ and/or \oii\ lines. The presence of these secondary samples is well known, including the fact that high redshift galaxies can be be misidentified for \ha\ emitters (and vice-versa). Line misidentification has already been pointed out by \citet{Pullen:2015yba} and more recently by \citet{Addison:2018xmc}.\footnote{In this respect, see also \citet{Gebhardt:2018zuj}, where they used the anisotropic power spectrum method \citep{Gong:2013xda} to estimate how the contaminated power spectrum changes for a given ratio of misidentified galaxies.} But misidentification will not happen for all high-$z$ galaxies and, in principle, one will be able to constitute samples of galaxies identifiable by other lines. In fact, WFIRST plans to constrain the baryon acoustic oscillations (BAO) scale in the redshift range $2<z<3$ using a sample of galaxies identifiable by their \oiii\ emission lines \citep{Spergel:2015sza}. \citet{Gebhardt:2018zuj} also consider the \oiii\ sample centred at $z=2.32$ contaminated by the low-$z$ \ha\ sample. \citet{Addison:2018xmc} took a similar approach for a \oiii\ sample centred in $z=1.9$ from a \euc-like survey. Although the last two works focus on the effects of line contamination, both of them neglect the potential contamination from \oii\ galaxies coming from even higher redshifts. 

But these works indicate the merit of looking for higher redshift star-forming ELGs using oxygen emission lines. Here we will take a step back and reinterpret these `interlopers' as an independent secondary galaxy samples, which we will use as a cosmological probe. We assume that one can clearly distinguish between emission lines. Indeed this discrimination between \oii, \oiiihb, and \ha\ can be possible using prior information from a sister photometric survey, as well as fainter lines such \hb\ in the observed spectra. Indeed, \citet{Pullen:2015yba} successfully demonstrated that using secondary lines and/or photometric information of the spectra one can separate the target sample from interlopers. Similarly, \citet{2013MNRAS.428.1498C} has used photometric information to construct a \oii\ sample at low redshift for SDSS, although for ELGs with a single emission line the success rate was lower. In addition, when two lines are present in the spectra, one can use prior knowledge of the line ratios \oii/\oiii\ and \oiii/\ha\ to assess which pair of lines is the most probable one. However, as \citet{Pullen:2015yba} pointed out, only a fraction of galaxies in a \ha\ sample will have secondary lines that can be used to avoid line misidentification (of order $\sim30$ to $50\%$ for WFIRST). Alternatively, one can rely on machinery such as the one developed by \citet{Kirby:2007yq} when only a single emission line is present. Some of these methods are dependent on the resolving capabilities of the spectrograph. Adding shape information of the line to the identification process will also be improve the classifiers. If we take the example of the \oiii\ doublet, both \euc\ and WFIRST would be able to resolve this double line (see Table \ref{tab:exp_details}). Hence, in light of the redshift ranges that each line can probe, one can ask if we can extend \euc\ and WFIRST (excluding SPHEREx) to cosmological probes of high-$z$ ELGs, and what is the merit of each individual sample for cosmology in the different redshift ranges. Although this possibility was known, only \citet{Pullen:2015yba} has studied the usage of \oii\ up to $z=2.38$ within the context of the Prime Focus Spectrograph \citep{Takada_2014}, and \oiii\ up to $z=2.9$ for WFIRST, as tracers of the large-scale cosmic structure at $z>2$. A possible explanation for this is the lack of available observationally calibrated luminosity functions at higher redshifts. Recent results from the High-$z$ Emission Line Survey \cite[HiZELS,][]{Geach:2008us} shed light on the redshift evolution of ELGs using the \oii\ and \oiiihb\ lines \citep{Khostovan:2015aa}. For recent semi-analytical works estimating the number of ELGs that would be seen using \ha\ and/or \oiii\ lines, see \citet{2019A&A...631A..82I} and \citet{Zhai:2019hjt}.

\begin{table*}
\centering            
\caption{Best-fit values for the \oiii+\hb\ and \oii\ luminosity functions from \citet{Khostovan:2015aa}.}
\begin{tabular}{cccccccccc}
\hline
&\multicolumn{4}{c}{\oiiihb}&&\multicolumn{4}{c}{\oii}\\
\cline{2-5}\cline{7-10}
Redshift & $0.84$ & $1.42$ & $2.23$ & $3.24$ && $1.47$ & $2.25$ & $3.34$ & $4.69$ \\
\hline\hline
$\log_{10}\phi_\ast$ & $-2.55^{+0.04}_{-0.03}$ & $-2.61^{+0.10}_{-0.09}$ & $-3.03^{+0.21}_{-0.26}$ & $-3.31^{+0.09}_{-0.26}$ && $-2.25^{+0.04}_{-0.04}$ & $-2.48^{+0.8}_{-0.09}$ & $-3.07^{+0.63}_{-0.70}$ & $-3.69^{+0.33}_{-0.29}$ \\
$\log_{10} L_\ast$ & $41.79^{+0.03}_{-0.05}$ & $42.06^{+0.06}_{-0.05}$ & $42.66^{+0.13}_{-0.13}$ & $42.83^{+0.19}_{-0.17}$ && $41.86^{+0.03}_{-0.03}$ & $42.34^{+0.04}_{-0.03}$ & $42.69^{+0.31}_{-0.23}$ & $42.93^{+0.18}_{-0.24}$ \\
$\alpha$ & $-1.6$ & $-1.6$ & $-1.6$ & $-1.6$ && $-1.3$ & $-1.3$ & $-1.3$ & $-1.3$ \\
\hline
\end{tabular}
\label{tab:LF_par}     
\end{table*}

These updated Schecter luminosity functions allow us to estimate the number density of observable high redshift objects for different flux thresholds. Based on these, we will compute the signal-to-noise ratio of the first multipoles of the power spectrum for different flux thresholds. Furthermore, we will assess and compare what kind of cosmological constraints one obtains from different survey areas and flux thresholds. We will show that the secondary high-$z$ samples complement the information we obtain from low-$z$ Universe, and present the case for them to be treated as independent cosmological samples. In fact, our results indicate that detailed studies of the precise number density estimations are needed, as well as development of machinery to disentangle the several galaxy samples. These are a requirement for proper calculations of the figure of merit of the secondary as a function of flux threshold and detection efficiency.

The paper is organised as follows: in \autoref{sec:nGal} we estimate the number of observable ELGs at different redshifts using simple prescriptions from observationally calibrated luminosity functions and in \autoref{sec:pk_multipole} we review the multipole expansion of the power spectrum. In \autoref{sec:detection} we present our main results such as signal-to-noise ratios for the high-z ELGs multipole power spectrum and forecasts of their constraining power. We finish in \autoref{sec:conclusions} discussion the feasibility and potential of the high redshift ELG sample.

In our analysis, we shall assume a {fiducial flat $\Lambda$CDM cosmology with parameters \citep{2018arXiv180706209P}}: $\Omega_{\rm m}=0.31$, for the present-day total matter fraction; $\Omega_{\rm b}=0.05$, for the present-day baryon fraction; $h\equiv H_0/(100\,\mathrm{km\,s^{-1}\,Mpc^{-1}})=0.6774$ for the dimensionless Hubble constant; and $n_{\rm s}=0.968$ and $A_{\rm s}=2.14\times10^{-9}$ for the slope and amplitude of the primordial power spectrum.

\begin{figure*}
    \centering
   \includegraphics[width=\textwidth]{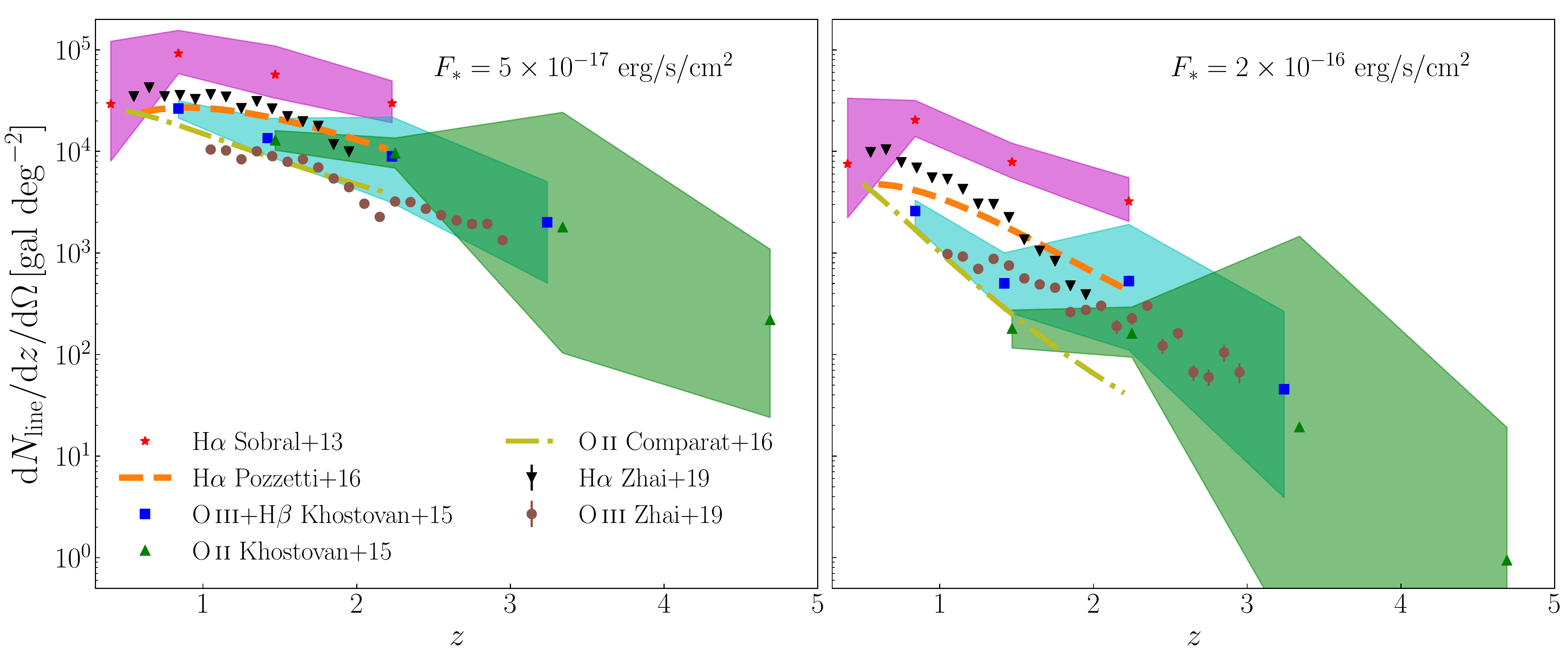}
    \caption{Estimates of the angular number density of ELGs as a function of redshift of the \ha\ sample, the \oiiihb\ sample, and the \oii\ sample, for a flux threshold of $F_\ast=5\times10^{-17}\,\mathrm{erg\,s^{-1}\,cm^{-2}}$ (left panel) and $F_\ast=2\times10^{-16}\,\mathrm{erg\,s^{-1}\,cm^{-2}}$ (right panel). The estimates were obtained from the Schecter \oiiihb\ and \oii\ luminosity functions of \citet{Khostovan:2015aa}, Schecter \ha\ luminosity functions of \citet{Sobral:2012fb}, modified Schecter \ha\ luminosity function (Model 3) of \citet{Pozzetti:2016cch}, Schecter \oii\ luminosity functions of \citet{Comparat:2016jqq}, and the results from simulations of \citet{Zhai:2019hjt} for \ha\ and \oiii.}
\label{fig:nELG_2Fs}
\end{figure*}

\section{The expected number of ELGs}
\label{sec:nGal}
As already discussed, emission lines other than \ha\ will be redshifted within the observable wavelength range of \ha\ galaxy surveys. The first natural approach to take is to consider each emission line as an individual sample. But those samples would have several overlapping galaxies and redshifts. For example, the spectrometers of \euc\ will work in the range $[0.92\,\mu\mathrm{m},1.85\,\mu\mathrm{m}]$ while WFIRST in the range $[1.0\,\mu\mathrm{m},1.93\,\mu\mathrm{m}]$ (with a second spectrograph at a lower range and lower resolution which details are yet to be finalised), as shown in \autoref{fig:z_vs_lamb}. Hence, it is more natural to break the ELG samples based on redshift ranges, rather than the line(s) used for the identification of the redshift of the host galaxy. We can, therefore, subdivide the foreseeable ELG samples into three redshift ranges:
\begin{itemize}
\item an ELG sample at $z\lesssim1.8$ using the \ha\ line in combination with other emission lines, which we call the \ha\ sample;
\item an ELG sample in the range $1.8\lesssim z\lesssim2.7$ using \oiii, \hb\, and \oii\ mainly, which we will call the \oiiihb\ sample;
\item an ELG sample at $2.7\lesssim z\lesssim4.0$ using \oii\ (alone or combined with other NUV lines), which we will call the \oii\ sample.
\end{itemize}
For the purpose of this paper, we will consider each sample independently and not a single ELG sample.

We will estimate the observed number density of \oiiihb\ and \oii\ galaxies using observationally calibrated Schecter luminosity functions which have the functional form, 
\be
\Phi(L)~\de\left(\frac{L}{L_\ast}\right)=\phi_\ast\left(\frac{L}{L_\ast}\right)^\alpha e^{-L/L_\ast}\de\left(\frac{L}{L_\ast}\right)\,.
\ee
The average comoving \textit{volumetric} density of a particular type of sources is given by
\be
\label{eq:dndv}
n_{\rm line}\,[\mathrm{gal\,Mpc^{-3}}] \equiv \frac{\de N_{\rm line}}{\de V} = \int_{L_{\rm min}/L_\ast}^{L_{\rm max}/L_\ast} \de \left(\frac{L}{L_\ast}\right) \Phi\left( L\right)\,,
\ee 
where the minimum luminosity is given by the flux threshold $F_\ast$, i.e.\ $L_{\rm min}(z)=4\pi\ D^2_{\rm L}(z)\ F_\ast$. $L_{\rm min}$ is redshift dependent via the luminosity distance is $D_{\rm L}(z)=(1+z)\chi(z)$, where $\chi(z)$ is the radial comoving distance. The maximum luminosity, $L_{\rm max}$, can formally be infinite, although in practice one cuts at a sufficiently large luminosity. This has little effect on the final estimate as the luminosity function is exponentially suppressed. Thus, the observed total \textit{surface} number of objects per steradian is given by
\be
\frac{\de N_{\rm line}}{\de z \, \de \Omega}\,[\mathrm{gal\,sr^{-1}}]= n_{\rm line}\, \frac{c\, D_{\rm A}^2}{H(z)}\,,
\ee
where the volume factor is given by the \textit{comoving} angular diameter distance $D_{\rm A}$ (which for a flat universe is the same as the comoving distance).

For the Schecter luminosity function one only requires a set of observationally calibrated parameters $\{\phi_\ast,\,L_\ast,\,\alpha\}$ for different lines/types of galaxies. In \autoref{tab:LF_par}, we summarise the results for the \oiiihb\ and the \oii\ samples found by \citet{Khostovan:2015aa} using HiZELS \citep{Geach:2008us}. In the right panel of \autoref{fig:nELG_2Fs}, we plot the estimates for the angular redshift distribution of \oiiihb\ and \oii\ sources for a experimental flux threshold of $F_\ast=2\times 10^{-16}\,\mathrm{erg\,s^{-1}\,cm^{-2}}$. The shaded areas represent the uncertainties in the number density from the luminosity function calibration errors. For comparison, we also include the estimates of the number of \oiii\ number of sources from \citet{Zhai:2019hjt} using simulations in combination with semi-analytical models. One can see that the expected numbers of \citet{Zhai:2019hjt} are within the shaded area given by \citet{Khostovan:2015aa}, although are systematically lower in the deep survey. For completeness, we also show the estimates for \ha: from \citet{Sobral:2012fb}, who calibrated a Schecter luminosity function using HiZELS; from \citet{Pozzetti:2016cch}, who also calibrated a modified Schecter luminosity function; from \citet{Comparat:2016jqq}, who compiled low redshift measurements of the number densities of \oii\ emitting galaxies; and the semi-analytical estimates of \citet{Zhai:2019hjt}. We present the estimates from the redshift-dependent luminosity function of \citet{Pozzetti:2016cch} and \citet{Comparat:2016jqq} up to the maximum redshift of the datasets included in their studies. As expected, there is a hierarchy of the number of ELGs detected at the same flux limit, as \oii\ is known to be weaker than \oiii, and the latter, in turn, weaker than \ha. In the left panel of \autoref{fig:nELG_2Fs}, we plot the expected numbers but for a flux threshold of $F_\ast=0.5\times10^{-16}\,\mathrm{erg\,s^{-1}\,cm^{-2}}$. Whilst the right panel of \autoref{fig:nELG_2Fs} can be regarded as the expected numbers for a wide survey such as \euc, the left panel of \autoref{fig:nELG_2Fs} can be understood as the expected numbers in a much deeper survey. This, in turn, corresponds to the volumetric number densities as shown in \autoref{fig:dNdV}, for several flux thresholds.
 \bfig
\centering
\includegraphics[width=\columnwidth]{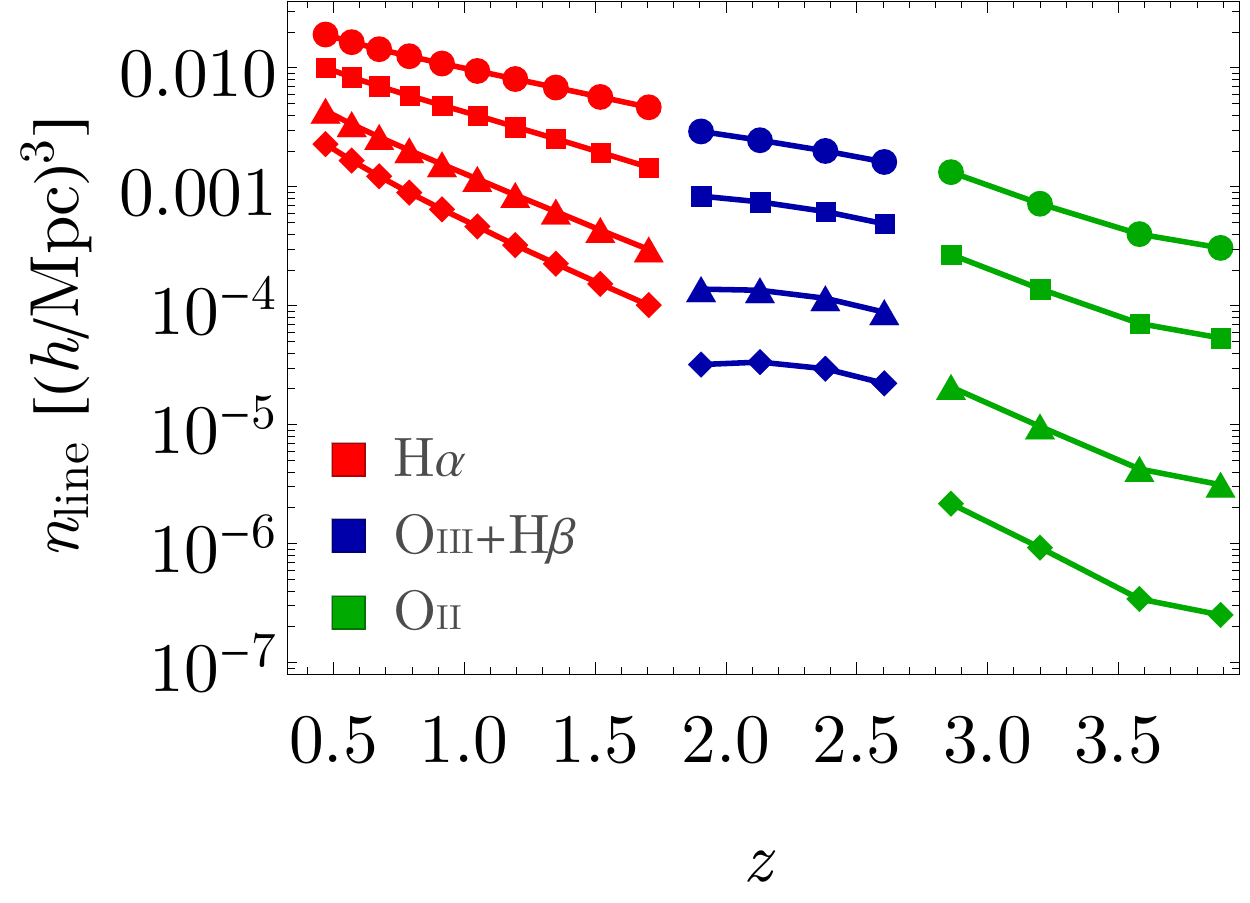}
\caption{Volumetric number densities of galaxies for the three ELG samples considered, as a function of flux threshold, ranging from $F_\ast=0.5\times10^{-16}\,\mathrm{erg\,s^{-1}\,cm^{-2}}$ (top) to $F_\ast=3\times10^{-16}\,\mathrm{erg\,s^{-1}\,cm^{-2}}$ (bottom).}
\label{fig:dNdV}
\efig

\section{O-line galaxy power spectrum and its multipoles}
\label{sec:pk_multipole}

For any biased tracer of the underlying cosmic large-scale structure, such are galaxies, the observed Fourier-space power spectrum of its number density fluctuations can be expressed as
\be
P_{\rm obs}(\bm k;z)=P(\bm k;z)+P_{\rm shot}(z)\,,\label{eq:Pobs}
\ee
for a wave-vector $\bm k$ at redshift $z$. Here, $P_{\rm shot}$ is a shot noise term and
\be
P(\bm k;z)=\left[b(z)+f(z)\mu^2\right]^2P_{\rm m}(k,z)\label{eq:Pk}
\ee
is the redshift-space galaxy power spectrum, with $b$ the linear galaxy bias (assumed to be scale-independent), $f$ the growth rate of density perturbations, $\mu$ being the cosine of the angle between the line-of-sight direction and the wave-vector $\bm k$, and $P_{\rm m}$ the power spectrum of matter density fluctuations, which only depends on $k=|\bm k|$ because of homogeneity and isotropy. In \autoref{eq:Pk}, the former term inside the square brackets is due to density fluctuations, and represents the dominant contribution, whereas the latter is the so-called redshift-space distortion (RSD) term. Lastly, $P_{\rm shot}$ is due to the Poisson nature of discretely sampling a continuous distribution, and it is thus simply given by the inverse of the volumetric number density of sources of \autoref{eq:dndv}, i.e.\
\be
P_{\rm shot}=\frac1{n_{\rm line}}\,.
\ee

\begin{table*}
\centering
\caption{Cumulative signal-to-noise ratio for the power spectrum multipoles, ${\rm SNR}_\ell$, for the various flux thresholds considered in the paper. Note that these numbers refer to full-sky measurements: to get the value corresponding to a survey covering $A_{\rm survey}$ steradians, it is sufficient to multiply the corresponding number by $[A_{\rm survey}/(4\pi)]^{1/2}$.}
\resizebox{\textwidth}{!}{
\begin{tabular}{cccccccccccc}
\hline
$F_\ast$ & \multicolumn{3}{c}{\ha} && \multicolumn{3}{c}{\oiiihb} && \multicolumn{3}{c}{\oii} \\
\cline{2-4}\cline{6-8}\cline{10-12}
$[\mathrm{erg\,cm^{-2}\,s^{-1}}]$ & ${\rm SNR}_{\ell=0}$ & ${\rm SNR}_{\ell=2}$ & ${\rm SNR}_{\ell=4}$ && ${\rm SNR}_{\ell=0}$ & ${\rm SNR}_{\ell=2}$ & ${\rm SNR}_{\ell=4}$ && ${\rm SNR}_{\ell=0}$ & ${\rm SNR}_{\ell=2}$ & ${\rm SNR}_{\ell=4}$ \\
\hline\hline
$0.5\times10^{-16}$ & $5.0\times10^3$ & $4.3\times10^3$ & $2.3\times10^3$ && $1.4\times10^3$ & $1.1\times10^3$ & $5.4\times10^2$ && $7.4\times10^1$ & $5.6\times10^1$ & $2.7\times10^1$ \\
$1.0\times10^{-16}$ & $4.1\times10^3$ & $2.5\times10^3$ & $6.8\times10^2$ && $1.1\times10^3$ & $6.4\times10^3$ & $1.6\times10^2$ && $6.3\times10^1$ & $3.5\times10^1$ & \colorbox{gray!25}{$8.3\times10^0$} \\
$2.0\times10^{-16}$ & $2.7\times10^3$ & $8.1\times10^2$ & $5.8\times10^1$ && $7.6\times10^2$ & $2.2\times10^2$ & $1.5\times10^1$ && $4.4\times10^1$ & $1.2\times10^1$ & \colorbox{gray}{{\color{white}$7.4\times10^{-1}$}} \\
$3.0\times10^{-16}$ & $1.9\times10^3$ & $2.5\times10^2$ & \colorbox{gray!25}{$5.7\times10^0$} && $5.3\times10^2$ & $6.9\times10^1$ & \colorbox{gray}{{\color{white}$1.4\times10^0$}} && $3.1\times10^1$ & \colorbox{gray}{{\color{white}$3.8\times10^0$}} & \colorbox{gray}{{\color{white}$7.2\times10^{-2}$}} \\
\hline
\end{tabular}}
\label{tab:SNR_l}
\end{table*}

For the rest of this analysis, we shall assume a common bias prescription \citep[see e.g.][for \ha\ galaxies]{Amendola:2012ys},
\be
b(z)=\sqrt{1+z},
\ee
since all the galaxies detected through the lines in consideration come from the same ELG sample. Despite this being a crude approximation, we emphasise that the exact value of the bias does not affect substantially the results we present. Moreover, the exact determination of the bias of the \oiiihb\ and \oii\ samples is beyond the scope of this paper.

Since RSDs induce an anisotropy in the power spectrum given by the dependence of $P$ on $\mu$, it is better to rewrite the observed galaxy power spectrum in a Legendre multipole expansion. Hence, we have
\be
P(\bm k;z)=\sum_\ell P_\ell(k;z)\ {\cal L}_\ell(\mu)\,,
\ee
where ${\cal L}_\ell(\mu)$ are the Legendre polynomials, and the coefficients $P_\ell(k)$ are uniquely dependent on the modules of the scale, $k$. 
The coefficients of the multipole expansion are then given by
\be
P_\ell(k;z)=\frac{2\ell+1}2\int_{-1}^1\de \mu\ P(\bm k;z) {\cal L}_\ell(\mu)\,.
\ee
Since $P(\bm k;z)$ is even in $\mu$, and the Legendre Polynomials have the same parity of its multipole index, only the even multipoles of the power spectrum are different from zero. It has been shown that the lowest multipoles carry the bulk of the cosmological information \citep[e.g.][]{Scoccimarro:2015bla,Bianchi:2015oia}. Therefore, we will only consider the first three non-zero multipoles, i.e.\ the monopole ($\ell=0$), the quadrupole ($\ell=2$), and the hexadecapole ($\ell=4$). It is easy to show that they read
\begin{align}
P_0(k;z)&=\left[b^2(z)+\frac23b(z)f(z)+\frac15f^2(z)\right]P_{\rm m}(k,z)\,,\\
P_2(k;z)&=\left[\frac13b(z)f(z)+\frac47 f^2(z)\right]P_{\rm m}(k,z)\,,\\
P_4(k;z)&=\frac8{35} f^2(z) P_{\rm m}(k,z)\,.
\end{align}

\section{Detectability of the signal}
\label{sec:detection}
Here, we explore the detectability of the cosmological signal at high redshift---namely $z\simeq2$ and beyond---described above.

\subsection{Signal-to-noise ratio}
In a given redshift bin $z_i$, we define the signal-to-noise ratio (SNR) for the power spectrum, neglecting RSDs, as
\be
{\rm SNR}(z_i)=\sqrt{\sum_j \left[\frac{P(k_j,\mu=0;z_i)}{\Delta P(k_j,\mu=0;z_i)}\right]^2}\,,
\ee
where the uncertainty on the measurement of a given mode is
\be
\Delta P (k_j,\mu;z_i)\simeq\sqrt{\frac{{2}}{N_k (k_j,z_i)}}P_{\rm obs}(k_j,\mu;z_i)\,.\label{eq:Delta_Pk}
\ee

\begin{figure}
\centering
\includegraphics[width=\columnwidth]{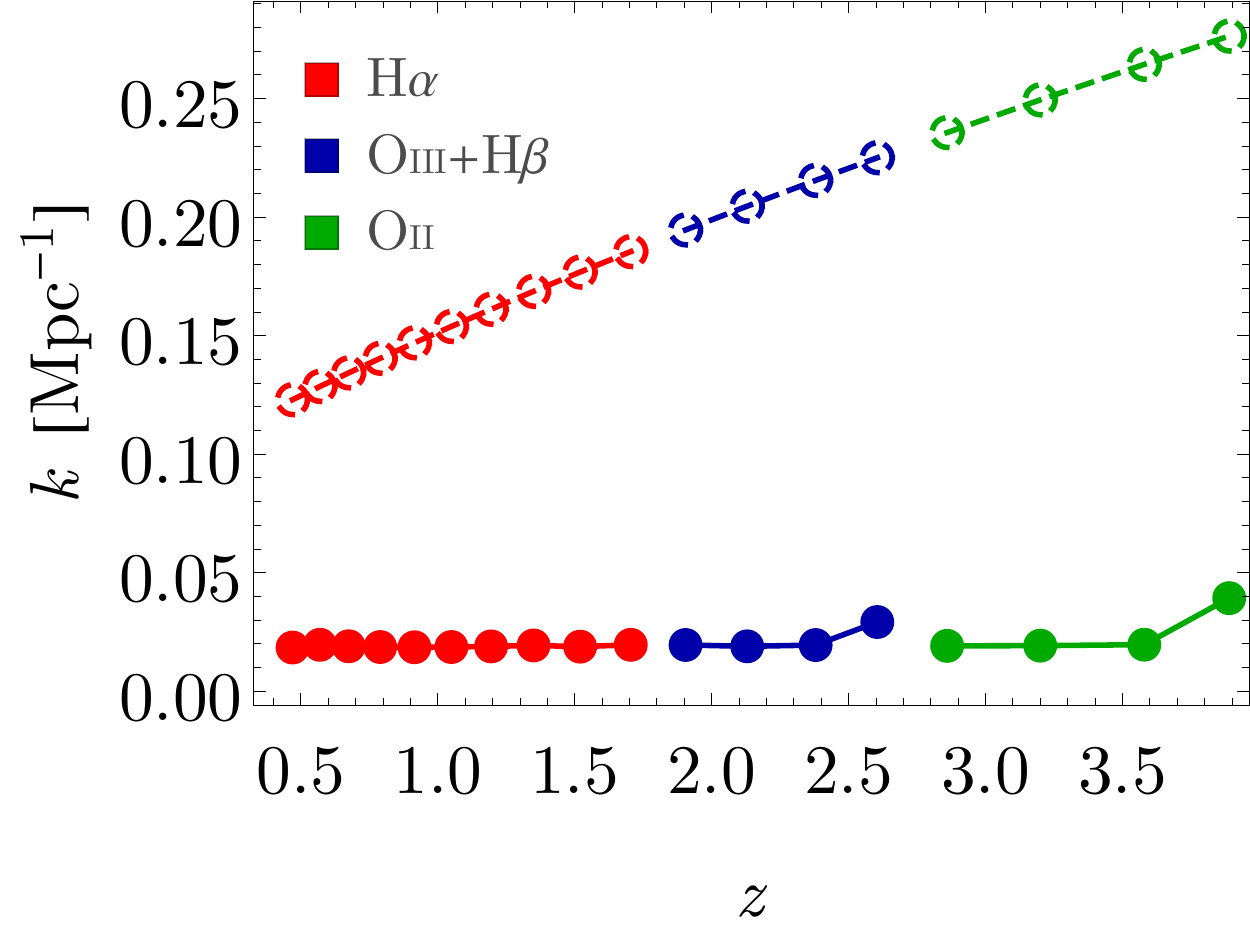}
\caption{$k_{\rm min}(z_i)$ (filled circles, solid lines) and $k_{\rm max}(z_i)$ (empty circles, dashed lines) in each redshift bin, for the three ELG samples. }\label{fig:k_range}
\end{figure}

The number of independent $k$-modes (omitting the redshift dependence) on a scale $k_j$, $N_k(k_j)$, depends on the volume of the survey. We follow the standard treatment and approximate it to be $N_k(k_j)\simeq k_j^2\Delta k V_{\rm survey}/(2\pi^2)$, where we take $\Delta k=k_{\rm min}\simeq 2\pi/L$, $L$ being the smallest side of the surveyed volume.\footnote{Note that another common choice in the literature is $k_{\rm min}\simeq 2\pi V_{\rm survey}^{-1/3}$, which, however, overestimates the constraining power on the largest scales for volumes that are not perfectly cubic.} Then, if follows that the sampled $k_j$'s go from $k_{\rm min}+\Delta k/2$ to (as close as possible to) $k_{\rm max}$, with $\Delta k$ as a step. We also stress that these quantities are all redshift-dependent, meaning that in fact we have $k_{\rm min}(z_i)$, $\Delta k(z_i)$ (the latter being equal to the former by construction), and $k_{\rm max}(z_i)$. Here we will take a very conservative approach to $k_{\rm max}(z_i)$ and truncate at linear scales. \citet{Smith:2002dz} has computed that the breakdown of linear matter perturbations happens at $k_{\rm nl}(z)=k_{\rm nl,0}(1+z)^{2/(2+n_{\rm s})}$, with $k_{\rm nl,0}\simeq 0.2\,h\,\mathrm{Mpc}^{-1}$. Therefore, at every redshift bin we will take $k_{\rm max}(z_i)=k_{\rm nl}(z_i)$. Neglecting small scales will lead to an underestimation the SNR, one should bear this in mind when reading the result presented in this paper. In \autoref{fig:k_range}, we show the minimum (filled circles, solid lines) and maximum (empty circles, dashed lines) wave-numbers as a function of redshift, for the three samples considered. The two `bumps' in $k_{\rm min}$ corresponding to the last redshift bins of both \ha\ and \oiiihb\ samples are due to the fact that those bins are slightly narrower that the others in each sample, to fit in each sample's wavelength range. 

To capture better the effect of RSDs, which induce an anisotropic pattern in the galaxy power spectrum, we also compute the SNR for Legendre multipoles, which reads
\be
{\rm SNR}_\ell(z_i)=\sqrt{\left[\sum_jP_\ell(k_j;z_i){\rm Cov}^{-1}_{\ell\ell^\prime}(k_j;z_i)P_{\ell^\prime}(k_j;z_i)\right]_{\ell=\ell^\prime}}\,,
\ee
where we have introduced the covariance of the $P_\ell$'s, viz.\
\begin{multline}
{\rm Cov}_{\ell\ell^\prime}(k;z)=\frac{2}{N_k(k,z)}\frac{(2\ell+1)(2\ell^\prime+1)}{2}\\
\times\int_{-1}^1\de\mu\,\left[{P_{\rm obs}}(k,\mu;z)\right]^2{\cal L}_\ell(\mu){\cal L}_{\ell^\prime}(\mu)\,,
\label{eq:Cov_multipoles}
\end{multline}
where the observed power spectrum is given in \autoref{eq:Pobs}. Note that in the Gaussian approximation we adopted, the multipole covariance is still diagonal in both redshifts and wave-numbers, but it is not in multipole. Finally, the total SNR, for either power spectrum or Legendre multipoles, is simply the sum in quadrature of the SNRs in each redshift bin.


\begin{figure}
\centering
\includegraphics[width=\columnwidth]{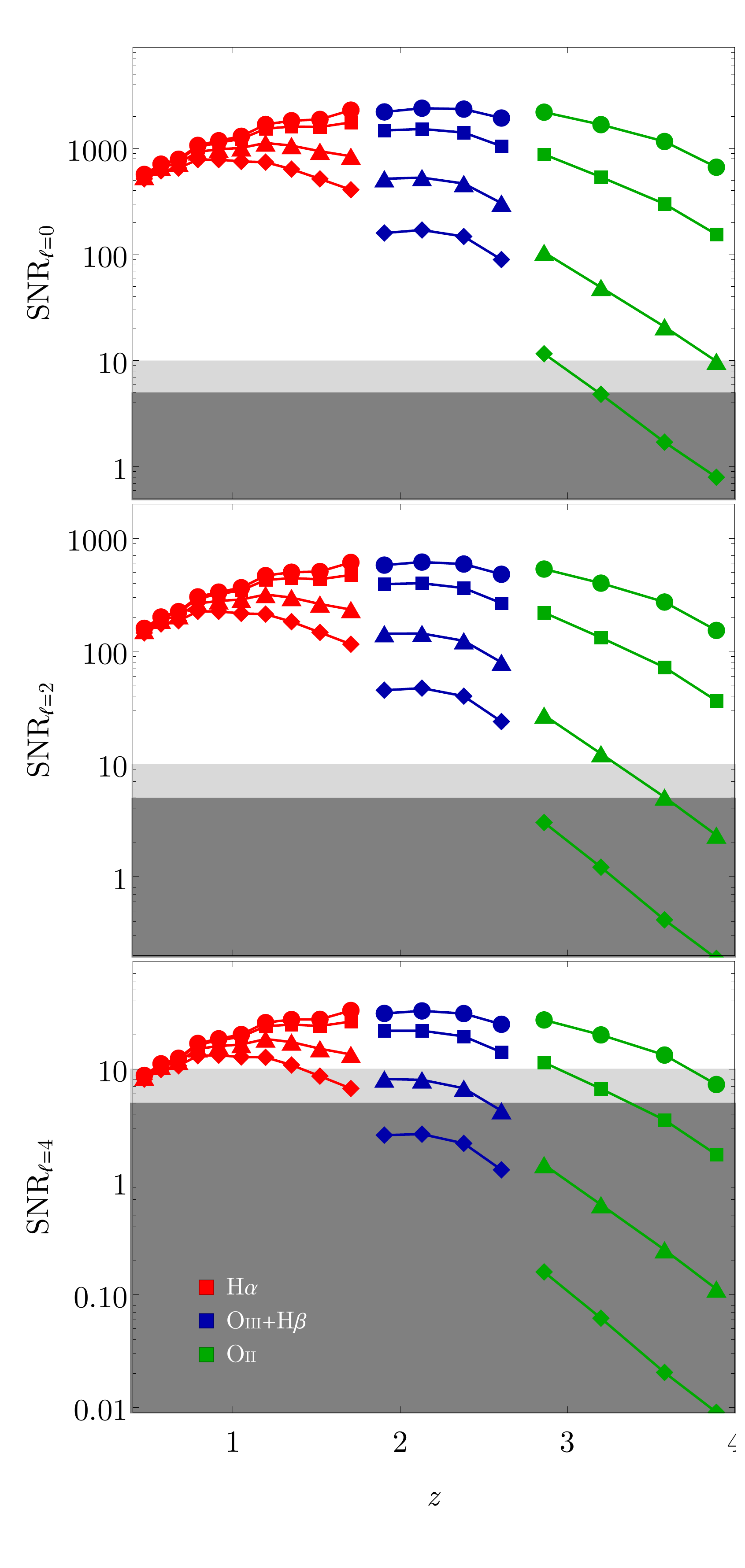}
\caption{${\rm SNR}_\ell(z_i)$ as a function of redshift for the first three Legendre multipoles of the galaxy power spectrum (red, green, and blue respectively for the \ha, \oiiihb, and \oii\ sample). Lines from top to bottom (and corresponding markers) refer to flux thresholds $F_\ast=(0.5,\,1.0,\,2.0,\,3.0)\times10^{-16}\,\mathrm{erg\,cm^{-2}\,s^{-1}}$. Dark- and light-grey areas denote regions of signal-to-noise ratios below $5$ and $10$, respectively. All curves refer to a hypothetical full-sky survey, and they should be rescaled by $0.6$ or $0.23$ in the case of \euc\ or WFIRST, respectively.}
\label{fig:SNR_l}
\end{figure}

\begin{figure*}
    \centering
    \includegraphics[width=\textwidth]{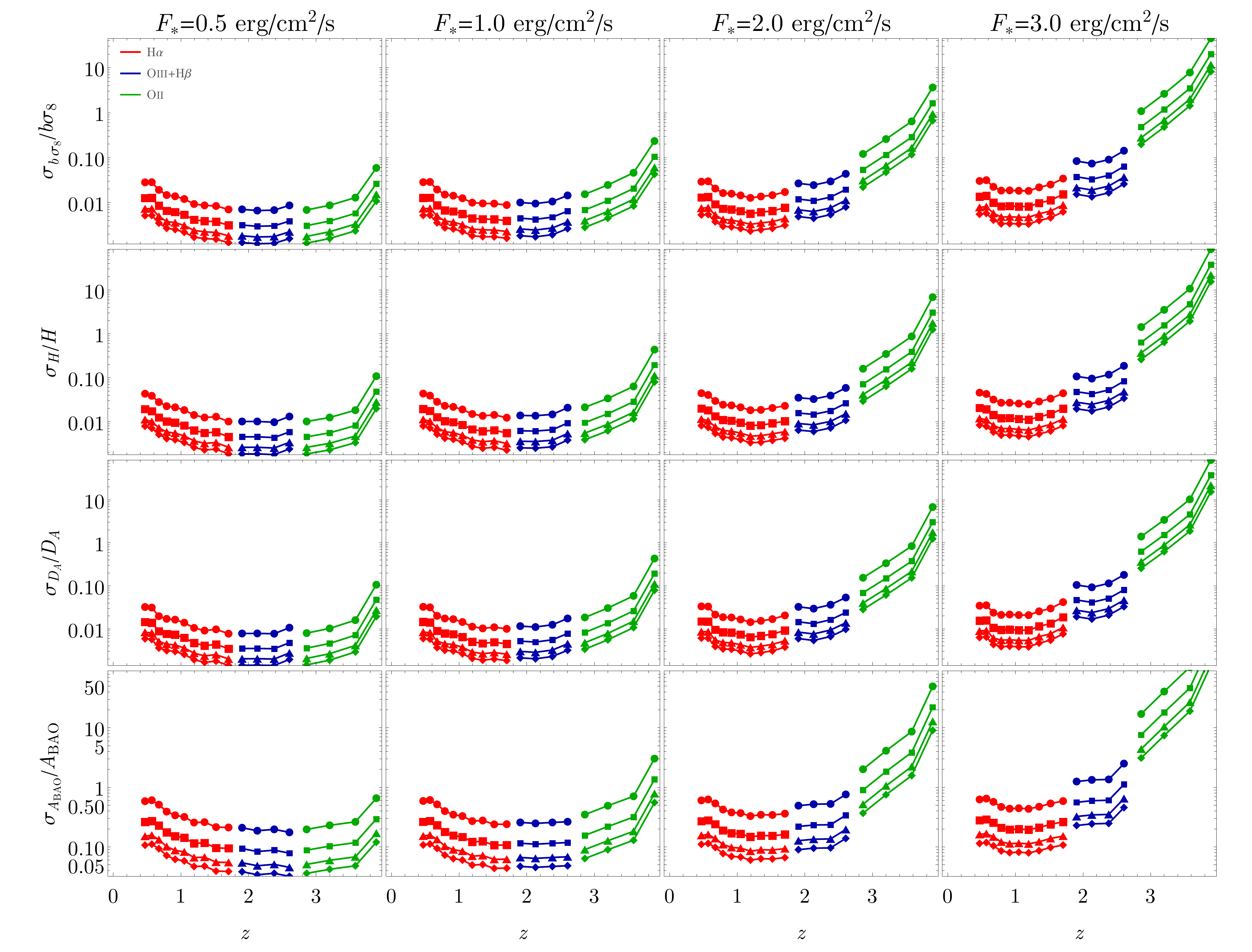}
    \caption{Relative $1\sigma$ marginal errors on redshift-dependent parameters $\{b\sigma_8(z),\,H(z),\,D_{\rm A}(z),\,A_{\rm BAO}(z)\}$ from the clustering of galaxies detected through different line emission: \ha\ in red, \oiiihb\ in green, and \oii\ in blue. Circles, squares, triangles, and diamonds respectively refer to survey areas of $1$, $5$, $15$, and $30$ in $10^3\,\deg^2$, whereas the four columns illustrate the dependence of the constraints on the flux threshold, set to $(0.5,\,1.0,\,2.0,\,3.0)\times10^{-16}\,\mathrm{erg\,cm^{-2}\,s^{-1}}$ from left to right.}
    \label{fig:rel_errs}
\end{figure*}

In \autoref{tab:SNR_l} we present the cumulative signal-to-noise ratio for the power spectrum multipoles, ${\rm SNR}_\ell$, for various flux thresholds and the three ELG samples considered in our analysis. For simplicity, we consider a full-sky survey and note that it is sufficient to rescale the numbers given in the table by the quantity $[A_{\rm survey}/(4\pi)]^{1/2}$, if one wants to know the cumulative ${\rm SNR}_\ell$ of a survey covering a sky area of $A_{\rm survey}$ steradians. This happens because the most relevant effect of a change in survey area is the rescaling (in the direction perpendicular to the line of sight) of $V_{\rm survey}$ in \autoref{eq:Cov_multipoles}---the third dimension, instead, is fixed by the redshift-bin width. Albeit it is true that when the transverse size of the survey volume becomes smaller than the radial one, the $k$-binning also changes because of the redefinition of $k_{\rm min}$ and, consequently, $\Delta k$; but this effect is largely subdominant compared to the overall linear dependence of ${\rm SNR}_\ell(z_i)$ upon $[A_{\rm survey}/(4\pi)]^{1/2}$.

As a take-home message from \autoref{tab:SNR_l}, we note all three Legendre multipoles will be in principle detectable at high significance (i.e.\ signal-to-noise ratio larger than $10$) even for the high-redshift \oiiihb\ and \oii\ samples. To guide the reader's eye, we highlight in the table in light/dark-grey the  pairs of flux thresholds and multipoles for which the cumulative signal-to-noise ratio is smaller than $5$ / falls within $5$ and $10$; in other words, those configurations in which the statistical power is insufficient / barely sufficient to detect the signal. In other words, we could be able to detect the monopole and the quadrupole of the galaxy power spectrum up to redshift $3-4$, extending significantly the reach of the \ha\ mother survey. This is further explored and clarified in \autoref{fig:SNR_l}, where the same full-sky but, this time, redshift-dependent ${\rm SNR}_\ell(z_i)$ is shown for the three main ELG samples. Panels from top to bottom respectively refer to the monopole, the quadrupole, and the hexadecapole. In each panel, line colours denote ELG samples (red for \ha, green for \oiiihb, and blue for \oii), and from top to bottom we show results for flux thresholds $F_\ast=0.5,\,1.0,\,2.0,$ and $3.0$ in units of $10^{-16}\,\mathrm{erg\,cm^{-2}\,s^{-1}}$. Light/dark-grey areas denote the regions of limited/no detection, viz.\ $5<{\rm SNR}_\ell(z_i)\le10$ and ${\rm SNR}_\ell(z_i)\le5$.


\begin{figure*}
    \centering
   \includegraphics[width=\textwidth]{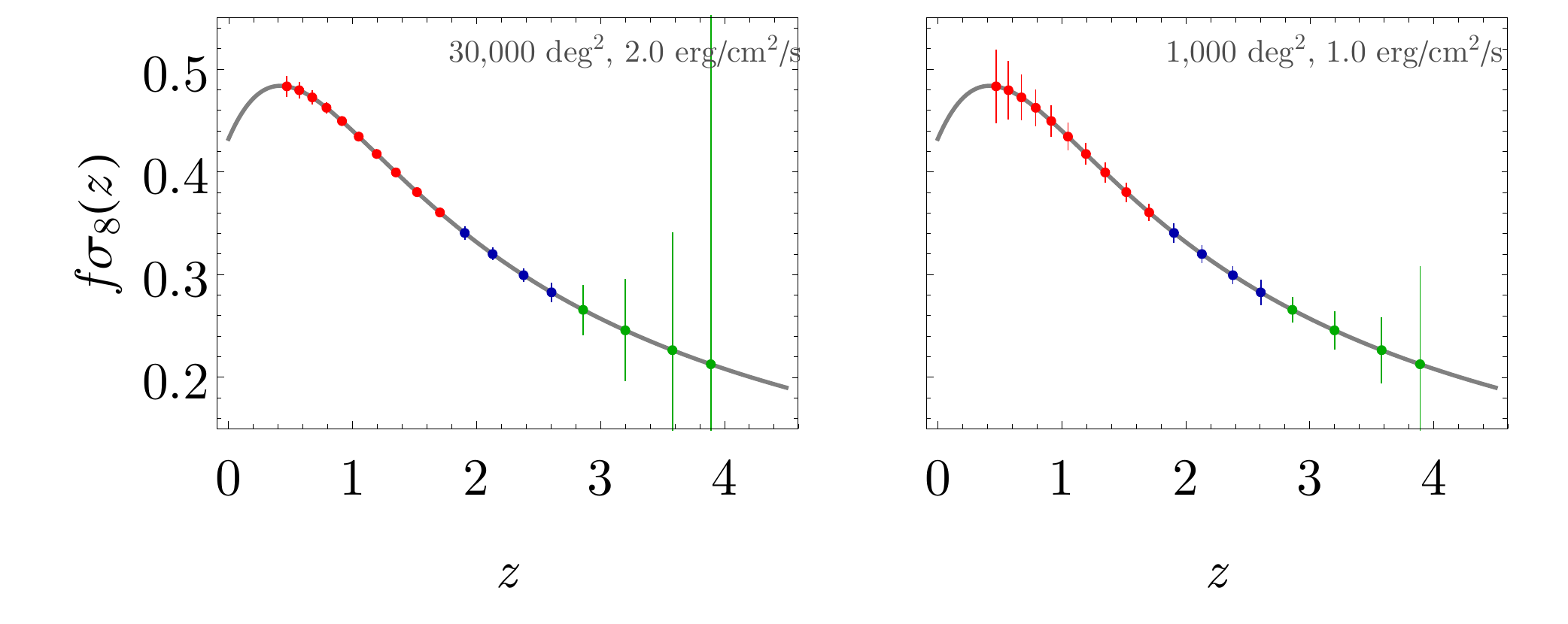}
    \caption{Relative $1\sigma$ marginal errors on  $f\sigma_8(z)$ from the clustering of galaxies detected through different line emission: \ha\ in red, \oiiihb\ in green, and \oii\ in blue, for two different surveys: on the left a lower sensitive but wide surveys ($30,000\,\deg^2$, $F_*=2\times10^{-16}\,\mathrm{erg\,s^{-1}\,cm^{-2}}$); on the right a narrow but more sensitive survey ($1000\,\deg^2$, $F_*=1\times10^{-16}\,\mathrm{erg\,s^{-1}\,cm^{-2}}$).}
    \label{fig:fs8}
\end{figure*}

\subsection{Estimation of cosmological parameters}
%

In the previous section, we have shown how the cosmological signal from \oiiihb\ and \oii\ galaxies is in principle detectable. Now, we move on and discuss its value for cosmological parameter estimation
. To do so, we will now consider five redshift-dependent cosmological parameters:
\begin{itemize}
    \item $A_{\rm BAO}(z_i)$, i.e.\ the amplitude of the baryon acoustic oscillation (BAO) `wiggles', defined as the oscillatory feature, $f_{\rm BAO}(k)$, on top of the smooth, broadband power spectrum, $P_{\rm smooth}(k)=P_{\rm m}(k)/[1+A_{\rm BAO}f_{\rm BAO}(k)]$;
    \item $b\sigma_8(z_i)\equiv f(z_i)D(z_i)\sigma_8$, i.e.\ the value of the linear galaxy bias multiplied by the square root of the overall normalisation of the matter power spectrum, $\sigma_8^2$, and the growth factor, $D(z)$;
    \item $f\sigma_8(z_i)\equiv b(z_i)D(z_i)\sigma_8$, i.e.\ the linear growth rate of structures, again factorising the redshift-dependent matter power spectrum normalisation;
    \item $H(z_i)$, i.e.\ the Hubble factor;
    \item $D_{\rm A}(z_i)$, i.e.\ the angular diameter distance.
\end{itemize}
We emphasise that each of the parameters described above is redshift dependent, meaning that we in fact constrain each of them separately in each redshift bin, centred on $z_i$.

A remark on the detection of BAOs is worth, at this point. The aforementioned approach may appear somewhat different to those most used in the literature, which focus either on locating the peak of the real-space two-point correlation function, or on the Alcock-Paczinsky test for the Fourier-space power spectrum \citep[e.g.][]{2018ApJ...863..110B,2018MNRAS.473.4773A}. Here, we instead focus on the detection of the BAO `wiggles' in terms of a non-zero amplitude of modulated $k$-features on top of an overall smooth, broadband spectrum. This is common in the literature involving Fisher matrix forecasts \citep{2008arXiv0810.0003R,Bull:2014rha}, and it allow us to treat the BAO amplitude parameter equally to the other parameters of the set. We emphasise that the scope of this paper is highlighting the added value of oxygen-line tracers to extend the reach of \ha\ surveys to high redshift; not presenting detailed forecasts for any specific cosmological mission.

The aforementioned parameters form a parameter vector $\bm\vartheta(z_i)$, for which we construct, in each redshift bin, a Fisher matrix according to
\be
F_{\alpha\beta}(z_i)=\frac12\int \de \mu\,\sum_j \frac{\partial_\alpha P(k_j,\mu;z_i)\partial_\beta P(k_j,\mu;z_i)}{\left[\Delta P(k_j,\mu;z_i)\right]^2}\,,
\ee
where $\partial_\alpha$ is a short-hand notation for the partial derivative taken with respect to $\vartheta_\alpha$. Hence, the cumulative Fisher matrix, $\mathbfss F$, is the sum of the $\mathbfss F(z_i)$ in each redshift bin. Then, the marginal error on a parameter $\vartheta_\alpha$ is given by 
\be
\sigma_{\vartheta_\alpha}=\sqrt{\left(\mathbfss F^{-1}\right)_{\alpha\alpha}}\,.
\ee

\autoref{fig:rel_errs} is a multi-panel plot summarising the relative marginal errors on parameters, $\sigma_{\vartheta_\alpha}/\vartheta_\alpha$, for all the parameters, the flux thresholds, the ELG samples and redshift bins, and the sky areas considered. In particular: each row refer to a specific parameters, namely $\{b\sigma_8(z_i),\,H(z_i),\,D_{\rm A}(z_i),\,A_{\rm BAO}(z_i)\}$ from top to bottom; each column refer to a specific flux threshold, i.e.\ $F_\ast=(0.5,\,1.0,\,2.0,\,3.0)\times10^{-16}\,\mathrm{erg\,cm^{-2}\,s^{-1}}$ from left to right; red, green, and blue lines respectively refer to \ha, \oiiihb, and \oii\ galaxies; and diamond, triangle, square, and circle markers refer to $(1,\,5,\,15,\,30)\times10^3\,\deg^2$, respectively.

The main conclusion one can draw from this plot is that---not counter-intuitively---sensitivity is possibly more important than area for high-redshift observations. This is further demonstrated in \autoref{fig:fs8}, where we focus on the extraction of RSDs in terms of constraints on the redshift-dependent quantity $f\sigma_8(z)$. We adopt the same colour code as before for the various ELG samples, and the two panels show forecast $1\sigma$ marginal error bars on measurements of $f\sigma_8(z_i)$ in each redshift bin, for a wide and shallow survey (left panel) or a narrow and deep survey (right panel). Clearly, measurements extracted from the original target, namely the \ha-galaxy sample, are optimised for the former survey specifications, with error bars $28-62\%$ tighter than those obtained with the latter experimental configuration. It turns out that a large area and a relatively larger flux threshold is also better for RSD estimation from the \oiiihb\ sample, with error bars $67-94\%$ smaller than for a narrow and deep survey. On the other hand, when it comes to the extraction of cosmological information from redshift $3-5$ \oii\ galaxies, it is better to observe as much as thirty times a smaller sky area, but with twice as deep a survey, which yields $f\sigma_8(z)$ measurements $25-38\%$ more constrained.

Finally, let us touch upon the importance of high-redshift oxygen-line galaxies to constrain cosmological parameters. Hitherto, we have focussed on model-independent parameters related to the redshift evolution of the matter power spectrum, and on its modulation by BAOs, galaxy bias, and RSDs. The next stage will now be to translate constraints on those parameters to more fundamental cosmological parameters as, for instance, the fractional densities of cold dark matter and baryons, or the equation-of-state parameters of dark energy---not to mention the possibility to study modifications to gravity at high redshift. All these choices are highly model-dependent and will give meaningful results only when applied to up-to-date survey specifications, we therefore defer such a scrutiny to a follow-up publication, involving realistic sky simulations, too.

\section{Discussion and conclusions}
\label{sec:conclusions}
We have generically used the locution `\ha\ surveys' for near-infrared space-based telescopes that will map galaxies positions in particular sections of the sky. Despite the abuse of terminology, the \ha\ line will take a prominent role in the spectroscopic determination of the redshift of a given detected galaxy. Although the prospects to extend \ha\ galaxy surveys up to $z\sim4$ are promising, we assumed that the samples can be identified unequivocally.

Although this is a strong assumption, we argue that it is possible to build statistically significant samples of oxygen lines from \euc\ and WFIRST. A full treatment of line identification is beyond the scope of this paper, but intuitively there are several ways to disentangle the contributions. For bright enough galaxies with several resolvable emission lines, misidentification will not be a problem as \citet{Pullen:2015yba} have shown.
 
Even when only one set of lines is visible, say \ha\ and \nii\ (or \oiiihb), then the line profiles will give an indication of which is the correct set. In the case of \oiiihb, the spectral resolution $R=380$, in combination with the equivalent width, may be enough to identify the \oiii\ doublet and the \hb\ line separately. 
Even if the \hb\ line turns out to be too faint, \euc's spectral resolution can separate the two lines on the \oiii\ doublet. 
As an example, an \oiii\ galaxy from $z=2.7$ will appear at the higher wavelength end of the range of \euc, where the resolution will be the worse. 
Still, \euc's resolution will be $\sim3.5$ times the redshifted separation of the two lines.  Another example of potential line confusion is when only a pair of strong lines are visible in the spectra. One might think that it would be \ha\ and \oiii, but using the pair separation, the equivalent width, and the ratio of the fluxes of the lines, one can in principle determine whether the pair corresponds to \ha\ and \oiii, or  \oiii\ and \oii\ (assuming that \hb\ is unresolvable). 
Additionally, the photometric sample can be used to identify the redshift of the galaxy, which in turn puts a strong prior on the line to be identified \citep{2013MNRAS.428.1498C}. Furthermore, it is possible to use the deep fields to train machine learning algorithms to identify lines even when a single faint line is present in the spectra \citep{Kirby:2007yq}. Hence, the combination of spectroscopic and photometric information together with line ratios, equivalent width and others can be used to train classifiers to construct the three different ELG samples proposed here.


Summarising, instead of removing higher redshift ELGs from the \ha\ sample, we propose to consider them as an entire new galaxy sample. It is therefore worth to use simulated spectra and assess how these different approaches can provide \oiiihb\ and \oii\ samples. On the other hand, if the confusion limit is too high, and the lines cannot be disentangled, one has to include the anisotropic power spectrum in the forward modelling and marginalise for the proportion of contamination \citep{Addison:2018xmc}, whilst fitting for the cosmological parameters.

One may ask what is the scientific merit for cosmology of these less numerous \ha\ contaminants. Therefore, we forecast how much information would the \oiiihb\ and \oii\ sample add to the standard set of cosmological parameters. We have shown that, despite worse constraining power than the low-$z$ \ha\ sample, the secondary high-$z$ samples can still provide percent level constraints on the expansion rate, growth, and the amplitude of the BAOs. As the Universe is more linear at higher redshifts, the reconstruction of the BAO is less demanding. Similarly, non-linearities only affect the power spectrum at scales smaller than in the late Universe. In addition to a tomographic study of the BAOs, a careful identification of the \oiiihb\ and \oii\ samples will allow for better tests of the growth and expansion rate up to a 1/5 of the size of the Universe. Current constraints from the high-$z$ post-epoch of reionisation Universe come mainly from the Lyman-alpha forest \citep[see e.g.][]{McDonald:2004eu} or its correlations with Quasars \citep[see e.g.][]{Font-Ribera:2013wce} or even its correlations with Damped Lyman-alpha systems \citep[see e.g.][]{Font_Ribera_2012}, although with less constraining power. While the \oiiihb\ sample can give similar constraints as the \ha\ sample, the \oii\ is very sensitive to the flux threshold of the experiment (as it quickly becomes shot-noise dominated). Even when the sample is noise dominated, the potentially large volumes allow for a statistical detection of the power spectrum. In the case of the \oii\ galaxy sample, we presented marginal errors without priors, but in fact we can put strong priors on $H_0$ and $\Omega_{\rm m}$ from other experiments (including the low redshift results from the same experiment), hence improving the constrains on $f\sigma_8$ at $z>3$. 

In this paper we asked ourselves the following question: \textit{given that \ha\ galaxy surveys can in principle observe higher redshift ELGs using other emission lines, is it possible to use those to obtain complementary cosmological constraints above $z>2$?} First we used recent state-of-the-art luminosity functions to estimated the number density of ELGs detectable using the \oiiihb\ set of lines and the \oii\ line. Despite the uncertainties inherited from the observationally calibrated luminosity functions and the fact that we assumed full observational efficiency, it seems possible to have enough detectable galaxies for a signal-dominated measurement. In fact, we saw in \autoref{fig:SNR_l} that the monopole cumulative signal-to-noise ratio is well above 5, if not even 10, for the three conceived samples, except for \oii\ in the faintest threshold limit. In \autoref{fig:rel_errs}, we showed the trade-offs between survey area and flux sensitivity, while for \oii\ is more sensitive to the flux threshold, \ha\ is more sensitive to the total sky area, as expected. Despite the technical details of future \ha\ surveys, it is worth to account and identify \oiiihb\ and \oii\ galaxies as they can increase substantially the overall of cosmological constraining power. More importantly, these 2 samples will work as an anchor between cosmic microwave background and local Universe constraints. It is therefore crucial to estimate properly the number densities of the secondary samples of \oiiihb\ and \oii, in order to have true signal-to-noise estimates and figures-of-merit for each survey.

\section*{Acknowledgements}
We thanks A. A. Khostovan for clarifications on the luminosity functions. We thanks L. Guzzo and B. Granett for useful discussions. JF is supported by the University of Padua under the STARS Grants programme CoGITO: Cosmology beyond Gaussianity, Inference, Theory and Observations. 
%
%
SC is supported by the Italian Ministry of Education, University and Research (\textsc{miur}) through Rita Levi Montalcini project `\textsc{prometheus} -- Probing and Relating Observables with Multi-wavelength Experiments To Help Enlightening the Universe's Structure', and by the `Departments of Excellence 2018-2022' Grant awarded by \textsc{miur} (L.\ 232/2016). 




\bibliographystyle{mnras}
\bibliography{Bibliography} 

\bsp 
\label{lastpage}
\end{document}